\documentclass[11pt,oneside,a4paper]{article}
\usepackage[T1]{fontenc}
\usepackage{newtxtext,newtxmath}
\usepackage{eurosym}
\usepackage{graphicx}
\usepackage{url}

\makeatletter
\g@addto@macro\bfseries{\boldmath}
\makeatother

\let\OLDthebibliography\thebibliography
\renewcommand\thebibliography[1]{
  \footnotesize
  \OLDthebibliography{#1}
  \setlength{\parskip}{0pt}
  \setlength{\itemsep}{0pt plus 0.3ex}
}

\newcommand{\epk}{\ensuremath{\epsilon_K}}
\newcommand{\eprk}{\ensuremath{\epsilon'_K}}
\newcommand{\Reps}{\ensuremath{{\rm Re}\,\epsilon'_K/\epsilon_K}}

\begin{document}
\pagestyle{empty}

\begin{flushright}
  {\bf KLEVER-PUB-18-02}\\
  {\bf December 18, 2018}
\end{flushright}

\vspace*{0.1\textheight}

\noindent\includegraphics[width=0.30\textwidth]{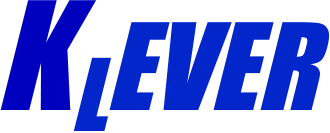}

\vspace{1.5ex}

{\bf \Large
  \noindent
  An experiment to measure ${\rm BR}(K_L\to\pi^0\nu\bar{\nu})$
  at the CERN SPS}

\vspace*{3.5ex}

\noindent\rule{\textwidth}{.4pt}

\vspace*{1ex}

{\bf \noindent Abstract}

\vspace*{1ex}

{\small
  \noindent Precise measurements of the branching ratios for the flavor-changing
  neutral current decays $K\to\pi\nu\bar{\nu}$ can provide unique constraints
  on CKM unitarity and, potentially, evidence for new physics. It is important
  to measure both decay modes, $K^+\to\pi^+\nu\bar{\nu}$ and
  $K_L\to\pi^0\nu\bar{\nu}$, since different new physics models affect the rates
  for each channel differently.
  The goal of the NA62 experiment at the CERN SPS
  is to measure the BR for the charged channel to within 10\%.
  For the neutral channel, the BR has never been measured.
  We are designing the KLEVER experiment to measure
  BR($K_L\to\pi^0\nu\bar{\nu}$) to $\sim$20\% using a high-energy neutral beam
  at the CERN SPS starting in Run 4.
  The boost from the high-energy beam facilitates the rejection of background
  channels such as $K_L\to\pi^0\pi^0$ by detection of the additional photons
  in the final state.
  On the other hand, the layout poses particular challenges for the
  design of the small-angle vetoes, which must reject photons from $K_L$
  decays escaping through the beam exit amidst an intense background from
  soft photons and neutrons in the beam. Background from $\Lambda \to n\pi^0$
  decays in the beam must also be kept under control.
  We present findings from our design studies for the beamline and experiment,
  with an emphasis on the challenges faced and the
  potential sensitivity for the measurement of
  BR($K_L\to\pi^0\nu\bar{\nu}$).}

\vspace*{1ex}

\noindent\rule{\textwidth}{.4pt}

\vspace*{5ex}

{\noindent\bf Input to the 2020 update of the European Strategy for
  Particle Physics} 

\vspace*{2ex}

\pagebreak

\noindent{\bf\Large The KLEVER Project}
\vspace*{2ex}

\newcommand{\fmark}[1]{\ensuremath{^{#1}}}

{\raggedright
\def\fNap{1}
\noindent
F.~Ambrosino,\fmark{\fNap}
\def\fRTV{2}
R.~Ammendola,\fmark{\fRTV}
\def\fFra{3}
A.~Antonelli,\fmark{\fFra}
\def\fGMU{4}
K.~Ayers,\fmark{\fGMU}
D.~Badoni,\fmark{\fRTV}
\def\fIns{5}
G.~Ballerini,\fmark{\fIns}
\def\fFer{6}
L.~Bandiera,\fmark{\fFer}
\def\fCEN{7}
J.~Bernhard,\fmark{\fCEN}
\def\fTur{8}
C.~Biino,\fmark{\fTur}
L.~Bomben,\fmark{\fIns}
V.~Bonaiuto,\fmark{\fRTV}
A.~Bradley,\fmark{\fGMU,a}
M.B.~Brunetti,\fmark{\fNap,b}
\def\fFlo{9}
F.~Bucci,\fmark{\fFlo}
A.~Cassese,\fmark{\fFlo}
R.~Camattari,\fmark{\fFer}
M.~Corvino,\fmark{\fNap}
\def\fPad{10}
D.~De~Salvador,\fmark{\fPad}
D.~Di~Filippo,\fmark{\fNap,c}
M.~van~Dijk,\fmark{\fCEN}
\def\fPis{11}
N.~Doble,\fmark{\fPis}
R.~Fantechi,\fmark{\fPis}
\def\fINR{12}
S.~Fedotov,\fmark{\fINR}
A.~Filippi,\fmark{\fTur}
\def\fMar{13}
F.~Fontana,\fmark{\fMar}
L.~Gatignon,\fmark{\fCEN}
\def\fSof{14}
G.~Georgiev,\fmark{\fSof}
A.~Gerbershagen,\fmark{\fCEN}
A.~Gianoli,\fmark{\fFer}
E.~Imbergamo,
\def\fPra{15}
K.~Kampf,\fmark{\fPra}
M.~Khabibullin,\fmark{\fINR}
\def\fPro{16}
S.~Kholodenko,\fmark{\fPro}
A.~Khotjantsev,\fmark{\fINR}
V.~Kozhuharov,\fmark{\fSof}
Y.~Kudenko,\fmark{\fINR}
V.~Kurochka,\fmark{\fINR}
G.~Lamanna,\fmark{\fPis}
M.~Lenti,\fmark{\fFlo}
L.~Litov,\fmark{\fSof}
E.~Lutsenko,\fmark{\fIns}
T.~Maiolino,\fmark{\fFer}
\def\fSNS{17}
I.~Mannelli,\fmark{\fSNS}
S.~Martellotti,\fmark{\fFra}
M.~Martini,\fmark{\fMar}
V.~Mascagna,\fmark{\fIns}
A.~Maslenkina,\fmark{\fINR} 
P.~Massarotti,\fmark{\fNap}
A.~Mazzolari,\fmark{\fFer}
E.~Menichetti,\fmark{\fTur}
O.~Mineev,\fmark{\fINR}
M.~Mirra,\fmark{\fNap}
M.~Moulson,\fmark{\fFra}
I.~Neri,\fmark{\fFer}
M.~Napolitano,\fmark{\fNap}
V.~Obraztsov,\fmark{\fPro}
A.~Ostankov,\fmark{\fPro}
G.~Paoluzzi,\fmark{\fRTV}
F.~Petrucci,\fmark{\fFer}
M.~Prest,\fmark{\fIns}
M.~Romagnoni,\fmark{\fFer}
M.~Rosenthal,\fmark{\fCEN}
P.~Rubin,\fmark{\fGMU}
A.~Salamon,\fmark{\fRTV}
G.~Salina,\fmark{\fRTV}
F.~Sargeni,\fmark{\fRTV}
V.~Semenov,\fmark{\fPro,\dagger}
A.~Shaykhiev,\fmark{\fINR}
A.~Smirnov,\fmark{\fINR}
M.~Soldani,\fmark{\fIns}
D.~Soldi,\fmark{\fTur}
M.~Sozzi,\fmark{\fPis}
V.~Sugoniaev,\fmark{\fPro}
A.~Sytov,\fmark{\fFer}
\def\fMIB{18}
E.~Vallazza,\fmark{\fMIB}
R.~Volpe,\fmark{\fFlo,d}
\def\fMai{19}
R.~Wanke,\fmark{\fMai}
N.~Yershov\fmark{\fINR}

\vspace*{2ex}
\noindent%
\fmark{\fNap}University and INFN Naples, Italy\\
\fmark{\fRTV}University and INFN Rome Tor Vergata, Italy\\
\fmark{\fFra}INFN Frascati National Laboratories, Italy\\
\fmark{\fGMU}George Mason University, Fairfax VA, USA\\
\fmark{\fIns}University of Insubria, Como, Italy\\
\fmark{\fFer}University and INFN Ferrara, Italy\\
\fmark{\fCEN}CERN EN-EA-LE, Geneva, Switzerland\\
\fmark{\fTur}University and INFN Turin, Italy\\
\fmark{\fFlo}University and INFN Florence, Italy\\
\fmark{\fPad}University of Padua and INFN Legnaro National Laboratories, Italy\\
\fmark{\fPis}University and INFN Pisa, Italy\\
\fmark{\fINR}Institute for Nuclear Research, Moscow, Russia\\
\fmark{\fMar}Marconi University and INFN Frascati National Laboratories, Italy\\
\fmark{\fSof}University of Sofia, Bulgaria\\
\fmark{\fPra}Charles University, Prague, Czech Republic\\
\fmark{\fPro}Institute for High Energy Physics, Protvino, Russia\\
\fmark{\fSNS}Scuola Normale Superiore and INFN Pisa, Italy\\
\fmark{\fMIB}INFN Milano Bicocca, Italy\\
\fmark{\fMai}University of Mainz, Germany\\
\vspace*{2ex}
\fmark{a}Now at Qrypt Inc., New York, USA\\ %Austin
\fmark{b}Now at University of Birmingham, UK\\ %Maria Brigida
\fmark{c}Now at Antaresvision SpA, Parma, Italy\\ %Mimmo
\fmark{d}Now at Universit\'e Catholique de Louvain, Belgium\\ %Roberta
\fmark{\dagger}Deceased\\
\vspace*{2ex}

\noindent Contact: Matthew Moulson, \url{moulson@lnf.infn.it}
}

\newpage
\setcounter{page}{1}
\pagestyle{plain}

\section{Scientific context}

The branching ratios (BRs) for the decays $K\to\pi\nu\bar{\nu}$
are among the observables in the quark-flavor sector most sensitive to
new physics. The Standard Model (SM) rates for these flavor-changing
neutral-current decays are very strongly suppressed by the GIM mechanism
and the CKM hierarchy.
In addition, because of the dominance of the diagrams with top loops,
the lack of contributions from intermediate photons, and the fact that 
the hadronic matrix element can be obtained from $K_{e3}$ data,
the SM rates can be
calculated very precisely: 
${\rm BR}(K^+\to\pi^+\nu\bar{\nu}) = (8.4\pm1.0)\times10^{-11}$ and 
${\rm BR}(K_L\to\pi^0\nu\bar{\nu}) = (3.4\pm0.6)\times10^{-11}$, where
the uncertainties are dominated by the external contributions from
$V_{cb}$ and $V_{ub}$ and the non-parametric theoretical uncertainties
are about 3.5\% and 1.5\%, respectively~\cite{Buras:2015qea}.

Because these decays are strongly suppressed and calculated very
precisely in the SM, their BRs are
potentially sensitive to
mass scales of hundreds of TeV, surpassing the sensitivity
of $B$ decays in most SM extensions~\cite{Buras:2014zga}.
Observations of lepton-flavor-universality-violating phenomena are mounting
in the $B$ sector~\cite{Bifani:2018zmi}.
Most explanations for such phenomena
predict strong third-generation couplings and thus significant changes
to the $K\to\pi\nu\bar{\nu}$ BRs through couplings to final states with
$\tau$ neutrinos~\cite{Bordone:2017lsy}.
Measurements of the $K\to\pi\nu\bar{\nu}$ BRs are critical
to interpreting the data from rare $B$ decays, and may demonstrate that
these effects are a manifestation of new degrees of freedom such as
leptoquarks~\cite{Buttazzo:2017ixm,Fajfer:2018bfj}.

The BR for the charged decay $K^+\to\pi^+\nu\bar{\nu}$ has been
measured by Brook\-ha\-ven experiment E787 and its successor, E949,
using $K^+$ decays at rest. The combined result
from the two generations of the experiment, obtained with seven candidate 
events, is ${\rm BR}(K^+\to\pi^+\nu\bar{\nu}) = 
1.73^{+1.15}_{-1.05}\times10^{-10}$ \cite{Artamonov:2009sz}.
The goal of the NA62 experiment at the CERN SPS
is to measure ${\rm BR}(K^+\to\pi^+\nu\bar{\nu})$ to within
10\%~\cite{NA62:2017rwk}.
NA62 has recently obtained the result
${\rm BR}(K^+\to\pi^+\nu\bar{\nu}) < 14\times10^{-10}$ (95\% CL),
with one observed candidate event, 0.267 expected signal events, and
$0.15^{+0.09}_{-0.03}$ expected background events \cite{CortinaGil:2018fkc}.
This result is based on 2\% of the combined data from running in
2016--2018. Additional running is contemplated for 2021--2022.

The BR for the decay $K_L\to\pi^0\nu\bar{\nu}$ has never been measured.
Up to small corrections, considerations of isospin symmetry lead to the
model-independent bound
$\Gamma(K_L\to\pi^0\nu\bar{\nu})/\Gamma(K^+\to\pi^+\nu\bar{\nu}) < 1$
\cite{Grossman:1997sk}.
Thus the 90\% CL limit ${\rm BR}(K^+\to\pi^+\nu\bar{\nu}) < 3.35\times10^{-10}$
from E787/E949 implies ${\rm BR}(K_L\to\pi^0\nu\bar{\nu}) < 1.46\times10^{-9}$.

Because the amplitude for $K^+\to\pi^+\nu\bar{\nu}$
has both real and imaginary parts, while the amplitude for
$K_L\to\pi^0\nu\bar{\nu}$ is
purely imaginary, the decays have different sensitivity to new
sources of $CP$ violation.
Measurements of both BRs would therefore be extremely useful not only
to uncover evidence of new physics in the quark-flavor sector, but also,
to distinguish between new physics models.
\begin{figure}[htb]
\centering
\includegraphics[width=0.5\textwidth]{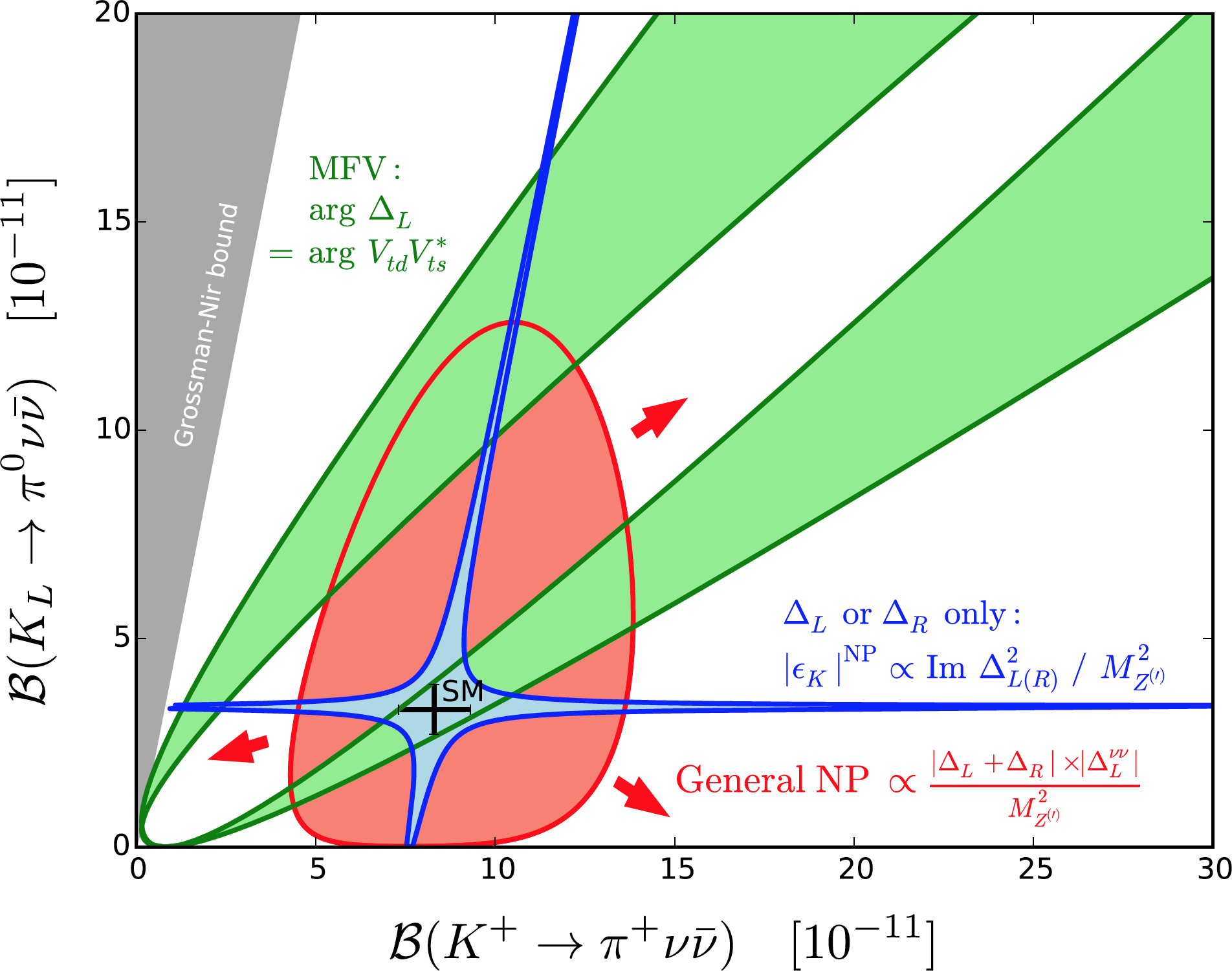}
\caption{Scheme for BSM modifications of $K\to\pi\nu\bar{\nu}$ BRs.} 
\label{fig:KpnnBSM}
\end{figure}
Fig.~\ref{fig:KpnnBSM}, reproduced from \cite{Buras:2015yca},
illustrates a general scheme for the expected
correlation between the $K^+$ and $K_L$ decays under various scenarios.
%\begin{itemize}
%  \item
    If the new physics has a CKM-like structure of flavor interactions,
    with only couplings to left-handed quark currents, the values for 
    ${\rm BR}(K^+\to\pi^+\nu\bar{\nu})$ and
    ${\rm BR}(K_L\to\pi^0\nu\bar{\nu})$
    will lie along the green band in the figure.
    Specifically, for new physics models with minimal flavor violation,
    constraints from other flavor observables limit the expected
    size of the deviations of the BRs to within about 10\% of their SM values
    (see, e.g., \cite{Isidori:2006qy}).
%  \item
    If the new physics contains only left-handed or right-handed couplings
    to the quark currents, it can affect $K\bar{K}$ mixing ($\Delta F = 2$)
    as well as $K\to\pi\nu\bar{\nu}$ decays ($\Delta F = 1$). Then, to
    respect constraints from the observed
    value of the $K\bar{K}$ mixing parameter $\epsilon_K$,
    the BRs for the $K^+$ and $K_L$ decays must lie on one of the
    two branches of the blue cross-shaped region \cite{Blanke:2009pq}.
    This is characteristic of littlest-Higgs
    models with $T$ parity \cite{Blanke:2009am}
    or in models with flavor-changing $Z$ or $Z'$ bosons and
    pure right-handed or left-handed
    couplings \cite{Buras:2014zga,Buras:2012jb}.
%  \item
    In the most general case, if the new physics has an arbitrary flavor
    structure and both left-handed and right-handed couplings, the
    constraint from  $\epsilon_K$ is evaded and there is little correlation,
    as illustrated by the red region.
    This is the case for the general MSSM framework (without minimal flavor
    violation) \cite{Isidori:2006qy} and in Randall-Sundrum models
    \cite{Blanke:2008yr}.
%\end{itemize}
  
Like ${\rm BR}(K_L\to\pi^0\nu\bar{\nu})$, the parameter \eprk\ also
gives a direct measurement of the height of the unitarity triangle.
Experimentally, \Reps\ has been measured by NA48 \cite{Batley:2002gn}
and KTeV \cite{Abouzaid:2010ny}, leading to the average
$\Reps = (1.66\pm0.23)\times10^{-3}$ \cite{Tanabashi:2018oca}.
In principle, the experimental value of \Reps\ significantly constrains
the value of ${\rm BR}(K_L\to\pi^0\nu\bar{\nu})$ to be expected in
any given new-physics scenario.
However, because of the delicate balance between the amplitudes
for the hadronic matrix elements of different operators, it is 
difficult to perform a reliable calculation of \Reps\ in the SM.
In a recent breakthrough, the RBC-UKQCD Collaboration obtained the
result $\Reps = (1.38\pm5.15\pm4.59)\times10^{-4}$
from a lattice calculation \cite{Bai:2015nea}, 2.1$\sigma$ less than
the experimental value
(improvements of the analytical parts of this result have increased the
significance slightly \cite{Buras:2015yba,Kitahara:2016nld}).
Considerations from large-$N_c$ dual QCD support the lattice
result \cite{Buras:2018wmb}, but a calculation in chiral perturbation
theory is in good
agreement with the experimental value of \Reps~\cite{Gisbert:2017vvj}.
RBC-UKQCD is currently working
on confirming the result and reducing both the statistical and systematic
uncertainties.

With this result for \Reps\ in the background, the correlation between
\eprk\ and ${\rm BR}(K_L\to\pi^0\nu\bar{\nu})$ has been examined in
various SM extensions at energy scales $\Lambda$ in the neighborhood of
1--10~TeV by several authors, in many cases with
constraints from \epk, $\Delta m_K$, and ${\rm BR}(K_L\to\mu\mu)$
considered as well.
The results of these studies are summarized in Table~\ref{tab:bsm}.
As a general rule, the observation of a larger value for \eprk\ than
expected in the SM implies a suppression of
${\rm BR}(K_L\to\pi^0\nu\bar{\nu})$ to below the SM value.
However, it is possible to construct models in which \eprk\ and
${\rm BR}(K_L\to\pi^0\nu\bar{\nu})$ are simultaneously enhanced.
With moderate parameter tuning (e.g., cancellation among
SM and NP interference terms to the 10-20\% level),
${\rm BR}(K_L\to\pi^0\nu\bar{\nu})$ may be enhanced by up to an order of
magnitude.
\begin{table}
  \centering
  {\scriptsize
  \begin{tabular}{lcccc}\hline\hline
    Model & $\Lambda$ [TeV] & Effect on ${\rm BR}(K^+\to\pi^+\nu\bar{\nu})$ & Effect on ${\rm BR}(K_L\to\pi^0\nu\bar{\nu})$ & Refs. \\ \hline
    Leptoquarks, most models & 1--20 & \multicolumn{2}{c}{Very large enhancements; mainly ruled out} & \cite{Bobeth:2017ecx} \\
    Leptoquarks, $U_1$ & 1--20 & +10\% to +60\% & +100\% to +800\% & \cite{Bobeth:2017ecx} \\
    Vector-like quarks & 1--10 & $-90$\% to +60\% & $-100$\% to +30\% & \cite{Bobeth:2016llm} \\
    Vector-like quarks + $Z'$ & 10 & $-80$\% to +400\% & $-100$\% to 0\% & \cite{Bobeth:2016llm} \\
    Simplified modified $Z$, no tuning & 1 & $-100\%$ to +80\% & $-100$\% to $-50$\% & \cite{Endo:2016tnu} \\
    General modified $Z$, cancellation to 20\% & 1 & $-100$\% to +400\% & $-100$\% to +500\% & \cite{Endo:2016tnu} \\
    SUSY, chargino $Z$ penguin & 4--6 TeV & & $-100$\% to $-40$\% & \cite{Endo:2016aws} \\
    SUSY, gluino $Z$ penguin & 3--5.5 TeV & 0\% to +60\% & $-20$\% to +60\% & \cite{Endo:2017ums} \\
    SUSY, gluino $Z$ penguin & 10 & Small effect & 0\% to +300\% & \cite{Tanimoto:2016yfy} \\
    SUSY, gluino box, tuning to 10\% & 1.5--3 & $\pm10$\% & $\pm20$\% & \cite{Crivellin:2017gks} \\
    LHT & 1 & $\pm20$\% & $-10$\% to $-100$\% &\cite{Blanke:2015wba} \\
    \hline\hline
  \end{tabular}
  }
  \caption{Effects on BRs for $K\to\pi\nu\bar{\nu}$ decays in various SM
    extensions, with constraints from other kaon observables, including
    in particular \Reps.}
  \label{tab:bsm}
\end{table}

The KOTO experiment at J-PARC is the only experiment currently pursuing
the decay $K_L\to\pi^0\nu\bar{\nu}$. 
KOTO continues in the tradition of the KEK experiment E391a in
technique \cite{Ahn:2007cd}.
The salient features of the experiment are the use of a highly collimated,
low-energy (mean momentum 2.1~GeV) ``pencil'' beam and very high performance
hermetic calorimetry and photon vetoing.  
KOTO has recently obtained the limit
${\rm BR}(K_L\to\pi^0\nu\bar{\nu}) < 3\times10^{-9}$ (90\% CL),
with an expected background of $0.42\pm0.18$ events and no candidate
events observed~\cite{KOTO:2018xxx}.
To reach the single-event sensitivity for the SM decay, KOTO would need
about a factor of 40 more data. With upgrades to the J-PARC Main Ring
to gradually increase the slow-extracted (SX) beam power to 100 kW, KOTO
should reach this threshold by the mid 2020s, and upgrades to the experiment
planned or in progress should reduce backgrounds by a similar
factor~\cite{KOTO:2018xxx}.
This would allow a 90\% CL upper limit on the BR to be set at the $10^{-10}$
level, but a next-generation experiment is needed in order to actually
measure the BR to test the
predictions in Table~\ref{tab:bsm}.

KOTO has long expressed a strong intention to upgrade to O(100)-event
sensitivity over the long term. To increase the kaon
flux, the production angle would be decreased from $16^\circ$ to $5^\circ$. 
This would require construction of a planned \$140M extension of the
J-PARC hadron hall, which is needed for other
experiments besides KOTO and is a priority for the Japan Science Council.
The mean $K_L$ momentum would be increased to 5.2~GeV. Since there are no
certain prospects to increase the SX beam power beyond 100 kW, a KOTO
step-2 upgrade would require the construction of a completely new
detector several times larger than the present detector (in part to
compensate for the effect of the increased beam momentum).
There is no official step-2 proposal, timeline, or sensitivity
estimate as of yet.

\section{Objectives of the KLEVER project}

We are designing the KLEVER experiment to use a high-energy neutral beam
at the CERN SPS to achieve a sensitivity of about 60 events for the decay
$K_L\to\pi^0\nu\bar{\nu}$ at the SM BR with an $S/B$ ratio of 1.
At the SM BR, this would correspond to a relative uncertainty of about 20\%,
demonstrating a discrepancy with $5\sigma$ significance if the true rate is
a bit more than twice or less than one-quarter of the SM rate,
or with $3\sigma$ significance if the true rate is less than half of the
SM rate.
These scenarios are consistent with the rates predicted for many different SM
extensions, as seen from Table~\ref{tab:bsm}.

\begin{figure}[htb]
  \centering
  \includegraphics[width=0.8\textwidth]{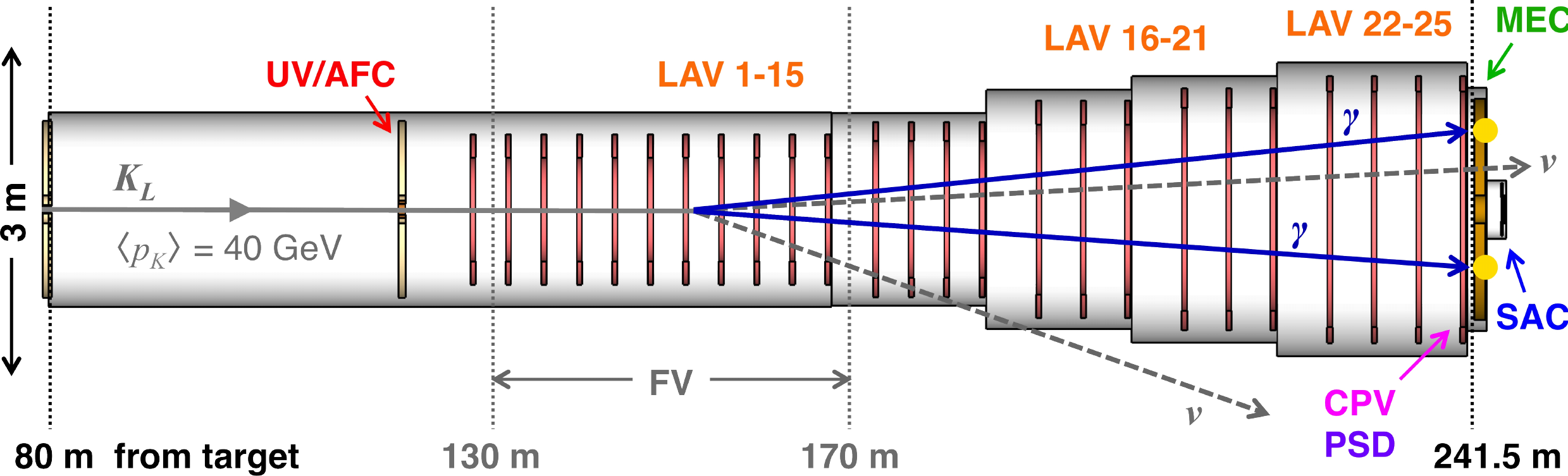}
  \caption{KLEVER experimental apparatus: upstream veto (UV) and
    active final collimator (AFC), large-angle photon vetoes (LAV), main
    electromagnetic calorimeter (MEC), small-angle calorimeter (SAC),
    charged particle veto (CPV), preshower detector (PSD).}
  \label{fig:exp}
\end{figure}
We do not envision KLEVER as a competitor to NA62, but rather as
a natural evolution of the multi-decade North Area program in kaon physics,
building on the successes of NA62 and its progenitors. The KLEVER timescale
is driven by the assumption that the kaon program will have advanced by the
end of LHC Run 3 to the point at which the measurement of
$K_L\to\pi^0\nu\bar{\nu}$ is the natural next step.
KLEVER would aim to start data taking in LHC Run 4 (2026).
The layout is sketched in Figure~\ref{fig:exp}. 
Relative to KOTO,
%which uses a neutral beam with a mean $K_L$ momentum of
%about 2~GeV,
the boost from the high-energy beam in KLEVER facilitates the rejection
of background channels such as $K_L\to\pi^0\pi^0$ by detection of the
additional photons in the final state.
On the other hand, the layout poses particular challenges for the
design of the small-angle vetoes, which must reject photons from $K_L$
decays escaping through the beam exit amidst an intense background from
soft photons and neutrons in the beam. Background from $\Lambda \to n\pi^0$
decays in the beam must also be kept under control.

\section{The KLEVER experiment}

\subsection{Primary and secondary beams}

KLEVER will make use of the 400-GeV SPS proton beam to produce a neutral
secondary beam at an angle of 8 mrad with an opening angle of 0.4 mrad.
The choice of production angle balances increased $K_L$ production and higher
$K_L$ momentum at small angle with improved $K_L/n$ and $K_L/\Lambda$ ratios
at larger angles. The choice of solid angle balances beam flux against
the need for tight collimation for increased $p_\perp$ constraints to reject
the main background from $K_L\to\pi^0\pi^0$ with lost
photons. The resulting neutral beam has
a mean $K_L$ momentum of 40 GeV, leading to an acceptance of
4\% for the fiducial volume extending from 130 to 170 m downstream of the
target, and a $K_L$ yield of $2\times10^{-5}$ $K_L$ per proton on target (pot).
With a selection
efficiency of 5\%, collection of 60 SM events would require a total
primary flux of $5\times10^{19}$ pot, corresponding to an intensity of
$2\times10^{13}$ protons per pulse (ppp) under NA62-like slow-extraction
conditions, with
a 16.8~s spill cycle and 100 effective days of running per year.
This is a six-fold increase in the primary intensity relative to NA62.

The feasibility of upgrades to the P42 beamline, TCC8 target gallery,
and ECN3 experimental cavern to handle this intensity has been studied
by the Conventional Beams working group in the context of the 
Physics Beyond Colliders initiative, and preliminary
indications are positive~\cite{Banerjee:2018xxx,Gatignon:2018xxx}:

\begin{itemize}
\item{\bf Slow extraction to T4}
The SPS Losses and Activation and PBC BDF/SHiP working groups have made
general progress on issues related to the slow extraction
of the needed intensity to the North Area, including duty cycle
optimization~\cite{Bartosik:2018xxx}.
An integrated intensity of $4\times10^{19}$ protons/year ($3\times10^{19}$
effective pot/yr after splitter losses) can be delivered to the TCC2 targets.
This would allow both KLEVER and a robust North Area fixed-target program
to run concurrently. With small compromises, SHiP, KLEVER, and the North
Area test beam program might be able to run concurrently, but
but it does not seem possible to run SHiP, KLEVER, and a robust
fixed-target program all at the same time without significant compromises. 

\item{\bf T4 to T10 transmission}
At the T4 target, the beam will be defocused and parallel in the vertical
plane, so that it mostly misses the 2 mm thick T4 target plate.
The small fraction that hits the target will be sufficient to produce the H6
and H8 beams without damaging the target head. On the other hand, most of the
beam will not be attenuated by T4 and the transmission to T10 could be as high
as 80\%. %This can be implemented for a 4.8-s flat top.

\item{\bf Target and dump collimator (TAX)}
Simulations show that the present T10 target cannot
withstand an intensity of $2\times10^{13}$ ppp. However, designs implementing
concepts used in the design of the CNGS target are capable of fulfilling
the requirements for the KLEVER beam. The simulations also show that
the K12 TAX is already close to thermal limits during running in beam
dump mode at the nominal NA62 intensity ($3\times10^{12}$ ppp).
A new design is needed with optimized block materials and
optimized cooling. This should be quite possible, as the TAX in the M2
beam has survived many years of operation with
$1.5\times10^{13}$ on the T6 target
The same arguments apply to the P42 TAX.

\item{\bf Containment of activated air}
There was initially a concern that the air containment in the TCC8
cavern would be insufficient for KLEVER operation with the existing
ventilation approach. However, the air flow inside the TCC8 target cavern
was measured to be very small, rendering an
expensive upgrade unnecessary.

\item{\bf Surface dose}
The primary proton beam is incident on the T10 target at downward angle
of 8 mrad and is further deflected downward by an additional sweeping
magnet following the target.
FLUKA simulations for the prompt dose on the surface above ECN3 are under
way; preliminary results indicate that an adequate solution has been found
for shielding the target region. Muons in the forward direction can
be dealt with by a mixed mitigation strategy involving
additional upstream shielding and potentially a thicker earthen shield
in the downstream region, possibly in combination with better fencing
around the ECN3 area.
\end{itemize}

\begin{figure}[htb]
  \centering
  \includegraphics[width=0.7\textwidth]{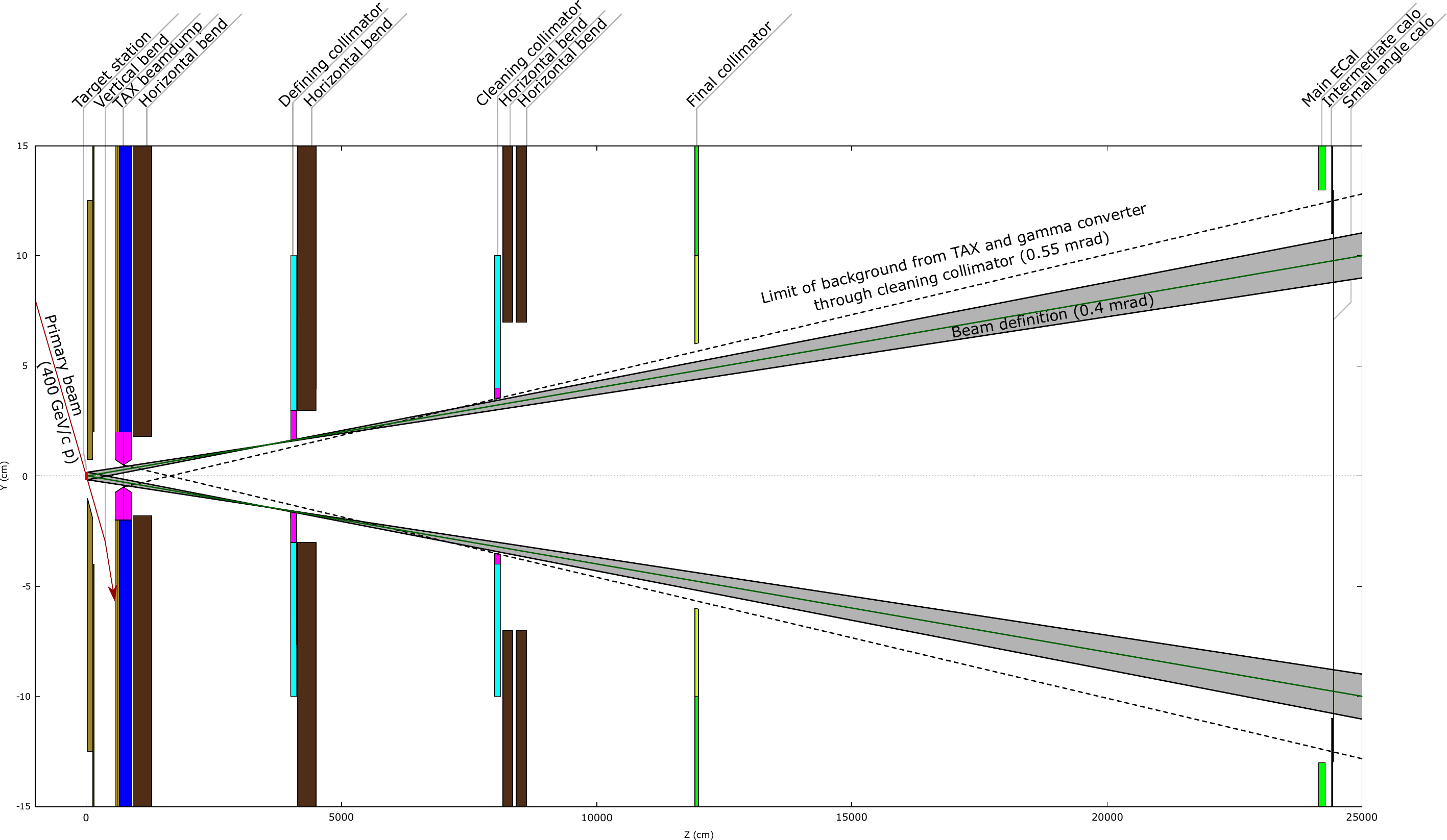}
  \caption{KLEVER neutral beamline layout as modeled in FLUKA,
    showing four collimation stages
    corresponding to dump, defining, cleaning, and final collimators.
    The final collimator is an active detector (AFC), built into the
    upstream veto. The aperture of the main calorimeter and coverage of the
    small-angle calorimeters is also shown.}
  \label{fig:beam}
\end{figure}

A four-collimator neutral beamline layout for ECN3 has been developed, as
illustrated in Fig.~\ref{fig:beam}. The primary proton beam, sloping
downwards at an angle of 8 mrad, is incident on the T10 target (assumed
to be a beryllium rod of 2~mm radius), producing the neutral
beam and other secondaries. This is immediately followed by a vertical
sweeping magnet that bends the protons further downward by 5.6 mrad,
and a TAX dump collimator with a hole for the passage of the neutral beam.
A horizontal sweeping magnet downstream of the TAX reduces the background
further.
A collimator at $z = 40$ m downstream of the target with $r = 16$~mm
defines the beam solid angle, and a cleaning collimator at $z = 80$~m
removes halo from particles interacting on the edges of the defining
collimator. Both are followed by horizontal sweeping magnets.
The sweeping fields have been optimized to minimize muon backgrounds.
The final stage of collimation is the active final collimator (AFC)
incorporated into the upstream veto (UV) at $z=120$~m. 
The cleaning and final (AFC) collimators have apertures which are
progressively larger than the beam acceptance, such that the former lies
outside a cone from the (2~mm radius) target passing through the defining
collimator, and the latter lies outside a cone from the TAX aperture
passing through the cleaning collimator.
The design of the beamline has been guided by FLUKA and Geant4 simulations
to quantify the extent and composition of beam halo, muon backgrounds,
and sweeping requirements.
According to the simulations, for an intensity of $2\times10^{13}$ ppp
and an effective spill length of 3~s, there are 140 MHz of $K_L$ in the beam
and 440 MHz of neutrons~\cite{vanDijk:2018xxx}.
An oriented tungsten crystal in the center of the TAX converts the vast
majority of the photons in the beam into $e^+e^-$ pairs, which are swept away,
but 40 MHz of photons with energy greater
than 5 GeV remain. As noted below, a crystal converter is used to maximize
the transparency to neutral hadrons at a given thickness in radiation
lengths (9.4$X_0$).

\subsection{Detector subsystems}

Because the experimental challenges involved in the measurement of
$K_L\to\pi^0\nu\bar{\nu}$, and in particular, the very high efficiency
required for the photon veto systems, 
most of the subdetector systems for KLEVER will have to be newly constructed.

\paragraph{Main electromagnetic calorimeter (MEC)}

Early studies indicated that the NA48 liquid-krypton calorimeter
(LKr) \cite{Fanti:2007vi}
currently used in NA62 could be reused as the MEC for reconstruction of
the $\pi^0$ for signal events and rejection of
events with additional photons. Indeed, the efficiency
and energy resolution of the
LKr appear to be satisfactory for KLEVER. However, the LKr time resolution
would be a major liability. The LKr would measure the event time in
KLEVER with 500 ps resolution, while the total rate of accidental vetoes
(dominated by the rate from beam photons interacting in the SAC)
could be 100 MHz.
The LKr time resolution might be improved via a comprehensive readout
upgrade, but concerns about the service life of the LKr would remain, and the
size of the inner bore would limit the beam solid angle and hence kaon flux.

We are investigating the possibility of replacing the
LKr with a shashlyk-based MEC patterned on the PANDA FS calorimeter (in turn,
based on the KOPIO calorimeter~\cite{Atoian:2007up}), with
$110\times110$~mm modules, lead absorber thickness of 0.275 mm, and scintillator
thickness of 1.5~mm, read out by silicon photomultipliers via 1-mm
diameter WLS fibers. Stochastic energy and time resolutions of better than
$\sigma_E/E = 2\%/\sqrt E$ and $\sigma_t = 72~{\rm ps}/\sqrt E$ were
obtained with this design in KOPIO tests.
We envisage a shashlyk design incorporating ``spy tiles'',
consisting of 10-mm thick scintillator bricks
incorporated into the shashlyk stack but optically isolated from it,
read out by separate WLS fibers. The spy tiles are located at key points
in the longitudinal shower development: near the front of the stack,
near shower maximum, and in the shower tail. 
This provides longitudinal sampling of the shower
development, resulting in additional information for $\gamma/n$ separation.
A first test of this concept was carried out with a prototype detector at
Protvino in April 2018.

\paragraph{Upstream veto (UV) and active final collimator (AFC)}

The upstream veto (UV) rejects $K_L\to\pi^0\pi^0$
decays in the 40~m upstream of the fiducial volume where there are
no large-angle photon vetoes. This is necessary, since it is possible
for $K_L\to\pi^0\pi^0$ decays in this region to place two photons 
on the fiducially accepted area of the MEC ($r > 35$~cm).
The UV is a shashlyk calorimeter with the same basic structure as the MEC
(without the spy tiles). 

The active final collimator (AFC) is inserted into a 100-mm hole in center
of the UV. The AFC is a LYSO collar counter with angled inner surfaces to
provide the last stage of beam collimation while vetoing photons from $K_L$
that decay in transit through the collimator itself. The collar is made of
24 crystals of trapezoidal cross section, forming a detector with an
inner radius of 60 mm.
The UV and AFC are both 800 mm in depth. The maximum crystal length for
a practical AFC design is about 250 mm, so the detector consists of 3 or
4 longitudinal segments. The crystals are
read out on the back side with two avalanche photodiodes (APDs).
These devices couple well with LYSO and offer high quantum efficiency,
simple signal and HV management. Studies indicate that a light yield
in excess of 4000 p.e./MeV should be easy to achieve.

\paragraph{Large-angle photon vetoes}

Because of the boost from the high-energy beam, it is sufficient for the
large-angle photon vetoes (LAVs) to cover polar angles out to 100 mrad.
The detectors themselves must have inefficiencies of less than
a few $10^{-4}$ down to at least 100 MeV, so the current NA62 LAVs based
on the OPAL lead glass cannot be reused.
The 25 new LAV detectors for KLEVER are
lead/scintillating-tile sampling calorimeters with wavelength-shifting fiber
readout, based on the CKM VVS design~\cite{Ramberg:2004en}.
Extensive experience with this
type of detector (including in prototype tests for NA62) demonstrates that
the low-energy photon detection efficiency will be sufficient for
KLEVER~\cite{Atiya:1992vh,Comfort:2005xx,Ambrosino:2007ss}.

\paragraph{Small-angle calorimeter}

The small-angle calorimeter (SAC) sits directly in the neutral beam and must
reject photons from $K_L$ decays that would otherwise escape via the
downstream beam exit. The veto efficiency required is not intrinsically
daunting (inefficiency $<1$\% for $5~{\rm GeV} < E_\gamma < 30~{\rm GeV}$
and $<10^{-4}$ for $E_\gamma > 30~{\rm GeV}$; the SAC can be blind for
$E_\gamma < 5~{\rm GeV}$),
but must be attained while maintaining insensitivity to more than 500~MHz
of neutral hadrons in the beam. In addition, the SAC must have good
longitudinal and transverse segmentation to provide $\gamma/n$
discrimination. In order to keep the false veto rate from accidental
coincidence of beam neutrons to an acceptable level ($<10$~MHz), and
assuming that about 25\% of the beam neutrons leave signals above threshold
in the SAC, topological information from the SAC must allow residual
neutron interactions to be identified with 90\% efficiency while maintaining
99\% detection efficiency for photons with
$5~{\rm GeV} < E_\gamma < 30~{\rm GeV}$. 

An intriguing possibility for the construction
of an instrument that is sensitive to photons and blind to hadrons is to
use a compact Si-W sampling calorimeter with crystalline tungsten tiles as
the absorber material, since coherent interactions of high-energy photons
with a crystal lattice can lead to increased rates of pair conversion
relative to those obtained with amorphous
materials~\cite{Bak:1988bq,Kimball:1985np,Baryshevsky:1989wm}.
The effect is dependent on photon
energy and incident angle; in the case of KLEVER, one might hope to decrease
the ratio $X_0/\lambda_{\rm int}$ by a factor of 2--3. The same effect could
be used to efficiently convert high-energy photons in the neutral beam to
$e^+e^-$ pairs at the TAX for subsequent sweeping, thereby
allowing the use of a thinner converter to minimize the scattering of
hadrons from the beam. Both concepts were tested in summer 2018 in the
SPS H2 beam line, in a joint effort together with the AXIAL collaboration.
In these tests, a beam of tagged photons with energies of up to 80 GeV was
obtained from a 120-GeV electron beam; interactions of the beam with
crystalline tungsten samples of 2 mm and 10 mm thick were studied as a
function of photon energy and angle of incidence with detectors just
downstream of the samples to measure charged multiplicity and forward
energy. While the data are still under analysis, the initial results show
a significant increase for both samples in the probability for
electromagnetic interactions when the crystal axis is aligned with the beam,
corresponding to the expected shortening of the radiation
length, with a somewhat smaller effect in the (lower-quality) thicker
crystal counterbalanced by a larger angular range over which the effect
is observed (a few mrad vs.\ about 1 mrad).

\paragraph{Charged-particle rejection}

For the rejection of charged particles, $K_{e3}$ is a benchmark channel
because of its large BR and because the final state electron can be mistaken
for a photon. Simulations indicate that the needed rejection can be achieved
with two planes of charged-particle veto (CPV) each providing 99.5\%
detection efficiency, supplemented by the $\mu^\pm$ and $\pi^\pm$ recognition
capabilities of the MEC (assumed in this case to be equal to those of the LKr)
and the current NA62 hadronic calorimeters and muon vetoes, which could be
reused in KLEVER. The CPVs are positioned $\sim$3~m upstream of the MEC
and are assumed to be constructed out of thin
scintillator tiles.
In thicker scintillation hodoscopes, the detection inefficiency arises
mainly from the gaps between scintillators. For KLEVER, the scintillators
will be only ~5 mm thick (1.2\% X0), and the design will be carefully
optimized to avoid insensitive gaps.

\paragraph{Preshower detector}

The PSD measures the directions for photons incident on the MEC.
Without the PSD, the $z$-position of the $\pi^0$ decay vertex can only be
reconstructed by assuming that two clusters on the MEC are indeed
photons from the decay of a single $\pi^0$. With the PSD, a vertex can be
reconstructed by projecting the photon trajectories to the beamline.
The invariant mass is then an independent quantity, and
$K_L\to\pi^0\pi^0$ decays with mispaired photons can be efficiently
rejected.
The vertex can be reconstructed using a single photon and the constraint
from the nominal beam axis. 
Simulations show that with $0.5 X_0$ of converter
(corresponding to a probability of at least one conversion of 50\%)
and two tracking planes with a spatial resolution of 100~$\mu$m,
placed 50 cm apart, the mass resolution is about 20~MeV and the
vertex position resolution is about 10~m. The tracking detectors
must cover a surface of about 5 m$^2$ with minimal material.
Micropattern gas detector (MPGD) technology seems perfectly suited for
the PSD. Information from the PSD will be used for bifurcation
studies of the background and for the selection of control samples,
as well as in signal selection.

\subsection{Simulation and performance}

Simulations of the experiment carried out with fast-simulation techniques
(idealized geometry, parameterized detector response, etc.) show that the
target sensitivity is achievable (60 SM events with $S/B = 1$). Background
channels considered at high simulation statistics include $K_L\to\pi^0\pi^0$
(including events with
reconstructed photons from different $\pi^0$s and events with overlapping
photons on the MEC), $K_L\to 3\pi^0$, and $K_L\to\gamma\gamma$.
\begin{figure}[htb]
  \centering
  \includegraphics[width=0.8\textwidth]{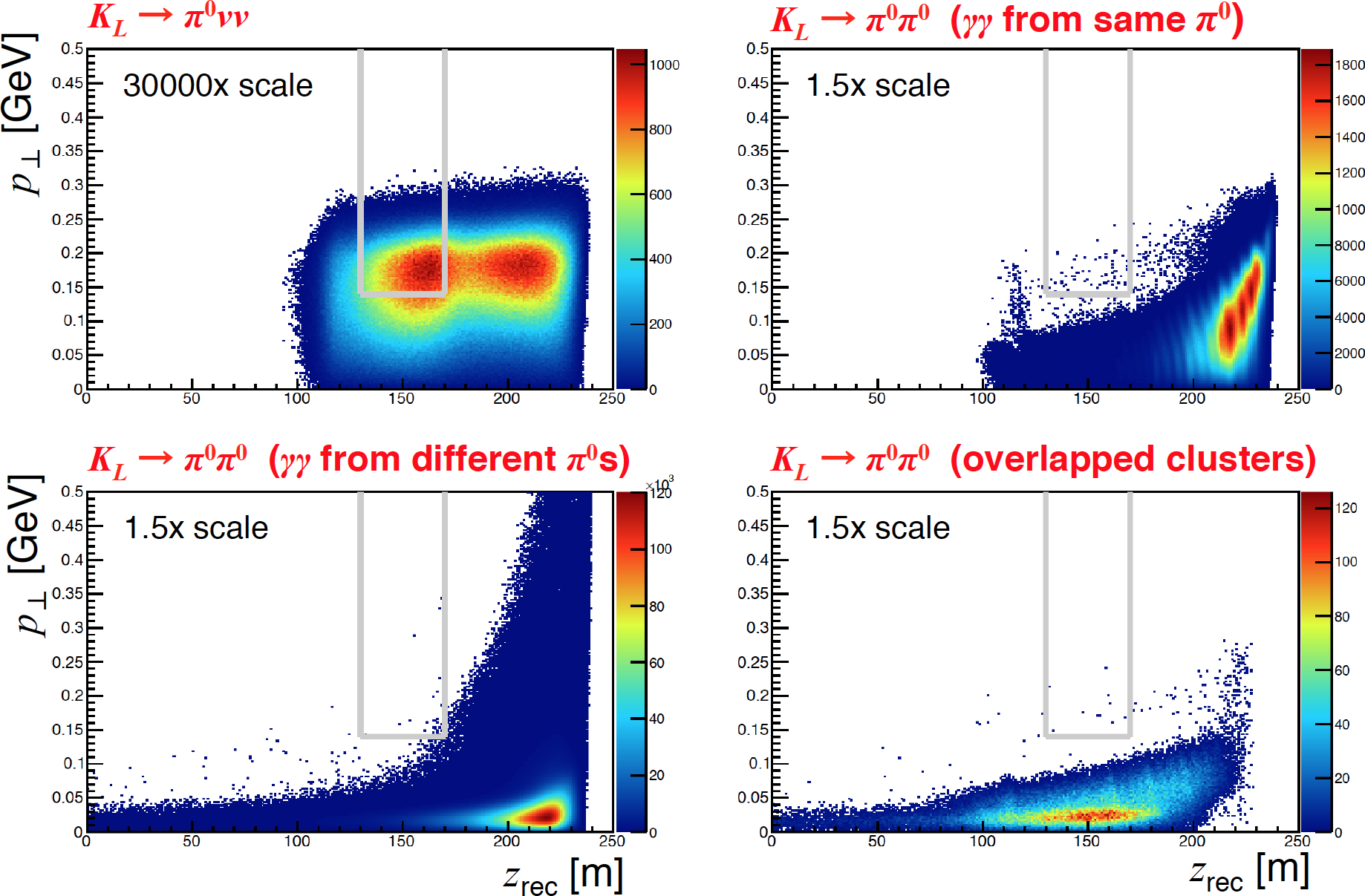}
  \caption{Distributions of events in plane of $(z_{\rm rec}, p_\perp)$
    after basic event selection cuts, from fast MC simulation, for 
    $K_L\to\pi^0\nu\bar{\nu}$ events (top left) and for
    $K_L\to\pi^0\pi^0$ events with two photons from the same
    $\pi^0$ (top right),
    two photons from different $\pi^0$s (bottom left), and
    with two or more indistinguishable overlapping photon clusters
    (bottom right).}
  \label{fig:sel}
\end{figure}
Fig.~\ref{fig:sel} illustrates the scheme for differentiating signal
events from $K_L\to\pi^0\pi^0$ background. Events with exactly two photons
on the MEC and no other activity in the detector are selected. The clusters
on the MEC from both photons must also be  more than 35~cm from the beam axis
(this helps to increase the rejection for events with overlapping clusters).
If one or both photons convert in the PSD, the reconstructed vertex
position must be inside the fiducial volume. The plots show the
distributions of the events satisfying these minimal criteria in the
plane of $p_\perp$ vs.\ $z_{\rm rec}$ for the $\pi^0$, where the
distance from the $\pi^0$ to the MEC is obtained from the transverse
separation of the two photon clusters, assuming that they come from a
$\pi^0$ ($M_{\gamma\gamma} = m_{\pi^0}$). This scheme is far from final and
there is room for improvement with a multivariate analysis, but is does
demonstrate that it should be possible to obtain $S/B\sim1$ with respect
to other $K_L$ decays.
Background
from $\Lambda\to n\pi^0$ and from decays with charged particles is assumed
to be eliminated on the basis of studies with more limited statistics.
This background is reduced by several orders of magnitude by the 130-m
length from the target to the fiducial volume and the choice
of production angle, which is carefully optimized to balance $K_L$ flux
against the need to keep the $\Lambda$ momentum spectrum soft.
Residual background from
$\Lambda\to n\pi^0$ ($p^* = 104$~MeV) can be effectively eliminated
by cuts on $p_\perp$ and in the $\theta$ vs. $p$ plane for the $\pi^0$.  

An effort is underway to develop a comprehensive simulation
and use it to validate the results obtained so far. Of particular note,
backgrounds from radiative $K_L$ decays, cascading hyperon decays, and
beam-gas interactions remain to be studied, and the neutral-beam halo
from our more detailed FLUKA simulations needs to be incorporated into
the simulation of the experiment.
While mitigation of potential background contributions from one or more
of these sources might ultimately require specific modifications to the
experimental setup, we expect this task to be straightforward in comparison
to the primary challenges from $K_L\to\pi^0\pi^0$ and $\Lambda\to n\pi^0$.

\subsection{Trigger, data acquisition, and readout}

Preliminary studies indicate that
the hit and event rates on most of the detectors are on the order of
a few tens of MHz, a few times larger than in NA62, with the notable
exception of the SAC, which will require an innovative readout solution
to handle rates of 100 MHz.
Digitization of the signals from the KLEVER detectors with FADCs at
high frequency (up to 1 GHz) would help to
efficiently veto background events without inducing dead time.
For the SAC, this is strictly necessary, since
the signal duration will be long compared to the mean interval between
events on a single channel. Detailed signal analysis may also assist with
particle identification and discrimination of uncorrelated background,
for example, from muon halo.

A free-running readout system would have various advantages:
there are no issues with latency, data are not buffered in the
front-end and thus not susceptible to radiation-induced corruption,
and above all, data selection is performed by software,
leading to a powerful and flexible trigger system that can in principle
implement the full offline analysis criteria, with no emulation required. 
On the other hand, a free-running, continuously digitizing readout implies
very challenging data rates. Assuming 100 MHz of interactions in the SAC
and current hit multiplicity and event size estimates, the data flow from
the SAC alone would be 100 GB/s.

A possible scheme for the readout system includes a low-level front-end
layer (L0) in which signals are digitized and timestamped using TDCs with
100 ps precision, and then transferred via a radiation hard link, for example,
via optical GBT link,
to a readout board such as the PCIe40 board under development for LHCb
or the FELIX board under development by ATLAS.
The first layer of data selection (L1) may be implemented on this
board. For example, methods for the $\gamma/n$ discrimination
in the SAC or cluster finding algorithms for the MEC may be implemented.
The output is written directly into the memory of the host PC, which
then routes the data to the online PC farm.
Once again, the data is received on a readout board for event building.
where the data may be preprocessed on GPUs or FPGA accelerators before
being loaded into the PC memory.
The specific solutions discussed here
are intended as examples to demonstrate that a readout system with the
needed requirements is within reach. Further evolution (including
KLEVER R\&D) is expected before actual solutions are chosen.

\section{Outlook}

KLEVER would aim to start data taking in LHC Run 4 (2026). Assuming a
delivered proton intensity of $10^{19}$ pot/yr, collection of 60 SM events
would require five years of data taking. To be ready for the 2026 start date,
detector construction would have to begin by 2021 and be ready for
installation by 2025, leaving three years from the present for design
consolidation and R\&D.
An Expression of Interest to the SPSC is in preparation.
%The timeline is discussed in further detail in the Addendum.

The KLEVER project builds on a CERN tradition of groundbreaking experiments
in kaon physics, following on the successes of NA31, NA48, NA48/2, and NA62.
Many institutes currently participating in NA62
have expressed support for and interest in the KLEVER project.
Successfully carrying out the KLEVER experimental program
will require the involvement of new institutions and groups, and we are
actively seeking to expand the collaboration at present.

\end{document}